\documentclass{kapproc}
\usepackage{t1enc}
\usepackage{procps}
\usepackage[dvips]{graphicx}
\upperandlowercase
\kluwerbib

\begin{document}

\articletitle{A statistical dynamical study of \\meteorite impactors:\\ a case study based on parameters\\ derived from the Bosumtwi impact event}
%\thanks{This paper includes 
%              data gathered with the Swope Telescope, located at Las Campanas 
%              Observatory, Chile}]

%\articlesubtitle{Photometry of 2060 Chiron (1977 UB) 
%                 and 10199 Chariklo (1997 $CU_{26}$)}

\author{M. A.  Galiazzo$^1$, \'A.  Bazs\'o$^1$, M. S. Huber$^2$, A. Losiak$^2$, R. Dvorak$^1$, C. Koeberl$^2$}

\affil{1. Institute for Astrophysics of the University of Vienna}%\\%
   % T{\"u}rkenschanzstra{\ss}e 17\\%
%}

\affil{2.  Department of Lithospheric Research, University of Vienna}
   % T{\"u}rkenschanzstra{\ss}e 17\\%

\email{%
    mattia.galiazzo@univie.ac.at}

% -------------------------------------

\begin{abstract}
   The study of meteorite craters on Earth provides information about the dynamic evolution of bodies within
the Solar System. Bosumtwi crater is a well studied, 10.5 km in diameter, ca. 1.07
Ma old   impact
structure located in Ghana. %(Koeberl et al., 1997a).
  The impactor was $\sim$1 km in
diameter, an ordinary chondrite 
%(Koeberl et al., 2007b)
 and  struck the Earth
with an angle  between $30^\circ$ and $45^\circ$  %(Artemieva et al., 2004)
  from the
horizontal. We have used a two phase backward integration to constrain the
most probable parent region of the impactor. We find that the most likely source
region is  a high inclination object from the Middle Main Belt.
\end{abstract}


% -------------------------------------

\section{Introduction}
 
When studying impact craters, it is sometimes possible to determine the
properties of the impactor that produced the crater, but the source where
the impactor originated in the Solar System is more difficult to determine.  Recently,
the Almahata Sitta fall was observed by astronomers, tracked by satellites as
it entered the atmosphere, and collected soon after striking Sudan.  In this
case, dynamical models were combined with detailed information about the
meteorite type to track the impactor back to the Inner Main Belt (Jenniskens et
al., 2010). For older impacts, the same precision cannot be achieved because
of the lack of detailed information on orbital parameters.  However, based
on the geological constraints on the dynamic nature of the impactor, a
statistical model can be used to suggest the most probable region from which the impactor
could have originated.  The aim of this study is to statistically constrain
the most probable parent region of the impactor that formed the Bosumtwi
impact crater.

\subsection{Bosumtwi crater: Geological background}
The Bosumtwi impact crater was chosen for this study because of its relatively
young age and unusually good constraints on the direction of the impactor. The
Bosumtwi impact crater is a 10.5 km in diameter complex meteorite impact
crater located in the Ashanti Province of southern Ghana. It is $1.07\pm0.11$ Ma old
and relatively well preserved (e.g., Koeberl et al., 1997a).  The Bosumtwi structure is
currently filled by the closed-basin Lake Bosumtwi that is 8 km in diameter
and up to 72.5 m deep. It is considered to be the largest, relatively
young, confirmed impact structure on the Earth.
Bosumtwi is a unique crater, since it is one of just three craters in the
world that are associated with a tektite strewn field (e.g., Koeberl,
1994).
 Tektites are centimeter-sized pieces of natural glass formed during a
hypervelocity impact event by ejection of molten target-surface material and
occurring in strewn fields (e.g., Koeberl, 1994). 
Based on the distribution of tektites around Bosumtwi crater it is possible to
constrain the direction of travel of the bolide prior to the impact.
 Based on the Cr isotope composition of the tektites derived from Bosumtwi, Koeberl et al. (2007b) established that the impactor that formed Bosumtwi crater was most probably an ordinary chondrite (while carbonaceous and enstatite chondrites were excluded).
The properties of the impactor that formed the crater have been constrained by
numerical modeling. According to Artemieva et al. (2004), the Bosumtwi structure was
formed by an impactor 0.75 to 1 km in diameter, moving with a velocity higher
than 15 km/s, and most probably 20 km/s. Due to association of the Bosumtwi
crater with the Ivory Coast tektite strewn field, the direction of the
incoming impactor was estimated to be from N-NE to S-SW and the angle of
impact is thought to be between  $30^\circ$ and $45^\circ$ (Artemieva et al., 2004).
% -------------------------------------

\section{Model \& Methods}
This study uses a statistical approach to constrain the parent region of the
Bosumtwi impactor, using $a-i$ space ($a$ and $i$
 for semi-major axis and orbital
 inclination, respectively) and the absolute magnitude ($H_v$)
 distribution inside the defined regions of the Solar System. 
 First, we made a backward integration\footnote{Due to the fact that a backward integration could be distorted by chaotic
motion in close encounters, we have looked for a measure that is
as simple as possible and is expressed in terms of orbital elements, since these are familiar indices of orbit differences. Because there
should be preferencial orbits in the regions far from the Earth, we can use
a statistical approach. We stopped the integration for a
particular body whenever the
asteroids overcome an eccentricity equal to 0.985 or have had a close encounter
with a planet less then $\sim10^{-5}$ AU.} from the
present to the time of impact. The integration used  the Radau integrator, included %full general
relativity, and all the planets plus Pluto, the Moon, Vesta, Ceres, Pallas and
Juno. The integration considered the positions of the Earth between 0.96 and 1.18
Ma ($1.07\pm0.11$ Ma) in order to find the possible position of the Earth during the time
when the impact occured, accounting for the error of the impact age measurement.
Then, we made another backward integration using the Lie-integrator
(Eggl and Dvorak, 2010)  without Mercury, Pluto and the 4 asteroids,
from the time of the impact to 100 Ma,
simulating the orbital evolutions of 924 fictitious Bosumtwi impactors
beginning at the calculated location of the Earth. Two cases were considered for this integration:
 
\begin{description}
\item [Fixed case (FC):] we started the integration at the location of the
  Earth (as calculated in the  initial integration) exactly at 1.07 Ma.
 Then, 384 particles, with a gaussian distribution of impact velocities ($v_{i}$) around 20
 km/s were launched with 32 different velocities. Those velocities correspond
 to the average value for Earth-impactors, as well
 as the most likely velocities indicated by numerical modeling for the
 Bosumtwi impactor (Artemieva et al., 2004). Velocities have a Gaussian distribution in the range of
 $11.2 $ to $40$ km/s, which are  the escape velocity from the Earth and 
 cometary speed, respectively. Then, 4 impact angles were considered using random values
 among $\Theta=37.5^\circ\pm7.5^\circ$ for each velocity and 3 different 
  directions ($\Omega_{1}=67.5\pm3.5^\circ$,
   $\Omega_{2}=78.75\pm3.5^\circ$ and $\Omega_{3}=56.25\pm3.5^\circ$ from east) for each
   angle.
 The launch position is the present latitude and longitude of the Bosumtwi crater site.
\item [General case (GC):] 540 particles were integrated using combinations
  of the following properties to account for the lack of knowledge of the
  exact position of the Earth at the time of the  impact:
 3 different  orbital positions  of the Earth, corresponding to the minimum, average
  and maximum aphelion (at 3 different times) in the Solar System; 3 different
  directions of the impactor ($\Omega_{1}=67.5\pm3.5^\circ$,
   $\Omega_{2}=78.75\pm3.5^\circ$ and $\Omega_{3}=56.25\pm3.5^\circ$ from east);
and 60 different sections of the Earth along lines of longitude every $6^\circ$
  for each position of the Earth. For each of the 540
  particles, impact angle and latitude\footnote{varying $\pm1.3^\circ$ (Neron deSurgy \& Laskar, 1995) from the present one, to account for the variance of the obliquity.} were distributed
  randomly, and $v_{i}$ had a gaussian distribution like in the FC.
\end{description}
%Regions are defined  in table \ref{region}. Groups are subdivided into
%high inclination  (\emph{HIG}, \citet{Nov2011}) 
%and low  inclination  (\emph{SIG}\footnote{% the osculatory
%borders of the most known family found by Zappal{\'a} et al. 3014 and we took
%, the highest limit, that it is for the Hilda Group (see also
%Broz 2008, for this group)}).
Once data were generated, analysis was done on two levels.  First, regions were
defined as in Table \ref{region}, where only the semimajor axis was
considered.  Then, for those particles which fell into the Main Belt, more specific
constraints were necessary because of the much higher population. Assuming the impactor was an ordinary
 chondrite (Koeberl et al., 2007b) and from the numerical results of Artemieva et al.\footnote{see also their
 Table 2 where the fit between the diameter
of the crater and the impact velocity is in agreement with impact velocities typical of asteroids.} (2004), we can
exclude the possibility of a cometary orbit, such as NEOs (Near-Earth objects) with orbits of
$Q>4.5$ AU (Fern{\'a}ndez et al., 2002). 

\begin{description}
\item[REGIONS  (Table \ref{region}):] At the end of the integration, the particles are examined to
  determine the probability that they fall into a defined region based on the semi-major axis
  range (called $P(a)$).  The orbital properties of the particles were derived
  from the time intervals between close encounters where they show little variance. The average time
 between close encounters with planets was
  determined to be 284 ky.

\item [MAIN BELT GROUPS  (Table \ref{MBAres}):] 
Asteroids in the Main Belt were subdivided into 3 regions 
  and with these 3 constraints: (1)  $1.5264<a<5.05$ AU, $Q<5.46$AU (aphelion of Hilda
  family from  Broz and Vokrouhlicky, 2008), (2) $q>1.0017$ AU (the average semi-major
  axis of the Earth after 100 Myr of integration)  and (3) the NEAs
  with $Q<4.35$ and $q>1.0302$. \\
 \noindent
Then, each of these 3  groups was divided into 2 subgroups: the low
inclination group (\emph{LIG}) and the high
inclination group (\emph{HIG}), the border between the two regions being
$i=17.16°$ (Novakovi{\'c} et al., 2011). The regions with the highest densities of
particles were then determined.  For these the Tisserand parameter with respect to
Jupiter: $T_j=\frac{a_j}{a}+2\sqrt{\frac{a}{a_j}(1-e^2)}\cos{i}$, where $a_j$ is the semi-major axis of Jupiter,
 $a$, $e$ and $i$ are the actions of the osculatory elements of the asteroid. It was calculated to test
whether or not the properties correspond to known families in the Main Belt.

\end{description}

\subsection{Absolute magnitude and spectroscopy}
Ordinary chondrites, thought to be responsible for the Bosumtwi impact,  are
associated with the  taxonomical S-group: S, L, A, K, R, Q and
intermediate types Sl, Sa, Sk, Sr, Sq (Bus \& Binzel, 2002).

Surveys have also revealed that the NEA population is dominated by  objects
belonging to the taxonomic classes $S$ and $Q$ (25\% as Q-type and 40 \% as
S-type, Bus et al., 2004).  When corrected for observational biases, about 40\% of the NEA population belong to one of these two taxonomic classes.
In the case of Mars crossers, 65\% belong to the S class (de L{\'e}on et al.,
2010). To compute the absolute magnitude of our impactor, we used the 
equation of Fowler and Chillemi (1992): $H_v=-5log(\frac{Dp^{1/2}}{1329})$.
Using the average albedo for the S-group asteroids, 0.197 (Pravec et al. 2012), and considering the likely size range of the Bosumtwi impactor, its
absolute magnitude ranged from 17.4 to 18.0 mag.
The absolute magnitude of the impactor can be used to
calculate the probability (called $P(H_V)$) that the impactor originated from a particular
region based on the likelihood of objects of similar absolute magnitude
originating in a particular
region, i.e., from the IMB, $11.21<H_{v_{IMB}}<27.60$. The spectral properties
of ordinary chondrites exclude the possibility that this object come from a
family such as Vesta or Hungaria, but it favours the  Flora, Ariadne, Nysa, Maria, Eunomia, Mersia, Walsonia,
Coelestina, Hellona, Agnia, Gefion and Koronis groups (Cellino et al., 2002) 
Barcelona and Hansa for the \emph{HIG} (Novakovi{\'c}, et al., 2011).

% -------------------------------------

\section{Results and discussion}
The final backward integration of 100 My shows that particles which survived the integration
tend to converge on the Main Belt (Figure \ref{one}),
and that only a negligible number of them is found in cometary orbits with an initial
aphelion greater than Jupiter's one, with a $v_{i}>27$ km/s; then
a very negligible part in hyperbolic orbits with a $v_{i}>33$ km/s.
 This suggests that the impactor most likely originated in the Main Belt.

\begin{description}
\item [REGIONS:]
The results are listed in Table 1, where $P(H_v)$ shows that on the basis of the
absolute magnitude, the object most likely originated from the Main Belt,
with a 37\% probability of originating from the IMB and a 29\% probability of
originating in the MMB.  The integration performed in this work shows that the majority of backwards
integrated particles fall into the IMB and MMB, with $\sim 10$\% of
objects in the FC falling into each of these.  The GC, however, resulted in
the majority of objects originating from the MMB, again with $\sim 10$\% of
objects, and only $\sim4$\% of objects originating from the IMB.

\begin{table}[!!h]
%\begin{minipage}{1.\textwidth}
\caption{Regions are defined by the $a$  that corresponds to strong 
perturbative Mean Motion Resonance (2.06
  AU for $J4:1$ and 3.28 for $J2:1$), apart for the inner border of the
  (\emph{IMB}) equal to the aphelion of Mars. The borders of the Jupiter
  Trojans (\emph{TRO}) as in  Tsiganis et al. (2005) and for the TNOs, the standard definition is
  used. \emph{MMB}, \emph{OMB},
  \emph{CEN} stand respectively for Middle Main Belt, Outer Main Belt and
  Centaurs. \emph{Orb.} stands for semi-major axis borders of the
  region. For the lower border of the IMB ($*$), we take the minimum $a$ for the innermost group
  of asteroids (see Galiazzo et. al. 2012).  The ``$\spadesuit$'' means that $H_v$
   are biased for the absence of a small bodies
  survey, so no significative computation is possible.
 $P_{FC}(a)$ and $P_{GC}(a)$ stands respectively for probability to find
  the origin in the region through the $a$ in the \emph{FC} and in
  the \emph{GC}.}
%\begin{center}
\begin{tabular*}{\textwidth}{@{\extracolsep{\fill}}lcccc}
\sphline
\sphline
\it Reg. & Orb. &
\it  $P(H_v)$ &
\it $P_{FC}(a)$ &
\it $P_{GC}(a)$  \\
\sphline
\sphline
IMB & $  1.78^\star \leq a \leq 2.06$    &0.3737 &0.0924 & 0.0404 \\
\sphline
MMB  &$2.06<a<3.28$       &0.2870 &0.1036  & 0.1030  \\
\sphline
OMB  &$3.28<a<5.05$      & 0.0232 &0.0112 &  0.0121   \\
\sphline
TRO&$5.05<a<5.35$   &$^\spadesuit$ &0.0056& 0.0000  \\
\sphline
CEN  & $5.35<a<30.00$    & $^\spadesuit$&0.056 & 0.0646      \\
\sphline
TNO  &$a>30.00$      &$^\spadesuit$ & 0.0112&  0.0020 \\
\sphline
\sphline
\end{tabular*}
\label{region}
%\end{minipage}
%\end{center}
\end{table}

\begin{table}[!!h]
%\begin{minipage}{1.\textwidth}
\caption{MBAs group have the same subdivision per semi-major axis, as in Table
  \ref{region}, apart for the IMB:
  $1.53<a<2.06$ where the lower limit is the average aphelion of Mars in
  100 Myr from the impact time). ``Low''$=$ Low inclined orbit ($i<17.16$) and
  ``High''$=$ High inclined orbit.$P_{F}(a,i)$ and $P_{G}(a,i)$ stands respectively for probability to find
  the origin in the region defined by semi-major axis, and inclination too, in the \emph{FC} and in
    the \emph{GC}.} 
\begin{tabular*}{\textwidth}{@{\extracolsep{\fill}}lcccc}
\sphline
\sphline
\it Reg. Low&
\it      $P_{F}(H_v)$&
\it  $P_{F}(a,i)$ &   
\it   $P_{G}(a,i)$  \\
\it Reg. High& 
\it     $P_{F}(H_v)$&
\it  $P_{F}(a,i)$ &  
\it    $P_{G}(a,i)$ \\
\sphline
\sphline
IMB  Low      & 0.234& 0.006 &0.000  \\
IMB High    &0.420&  0.006  & 0.018\\
\sphline
MMB   Low&  0.307&0.000     &0.010   \\
MMB  High& 0.113 &0.020       &0.022  \\
\sphline
OMB   Low     & 0.023& 0.003  &0.000   \\
OMB  High&  0.019 &0.000    & 0.004  \\
\sphline
\sphline
\end{tabular*}
\label{MBAres}
%\end{minipage}
\end{table}

\item[MAIN BELT:]
Results are  given in Table \ref{MBAres}, subdivided in 2 rows. In the upper
row we have the \emph{LIG}
and the lower one, the \emph{HIG}: $P(a,i)$ is the probability to find the
asteroid at high or low inclinations in the regions defined via semi-major
axis, $G$ and $F$ stands respectively for GC and FC.

\end{description}

%\clearpage

\begin{description}
\item [FC:] The most probable source region of the Bosumtwi impactor based on the
  fixed case integration falls within the Main Belt at high inclination, with the most
  likely group being
  the MMB at high inclination, with 2\% of the population falling into this
  group.  The objects have highly inclined orbits (up to $\sim75^\circ$), and the
  most populated zone at  $2.42\pm0.03<T_j<2.84\pm0.25$.

\item [GC:]  The most probable source region of the Bosumtwi impactor based on the
  general case is from the MMB with high inclination: $i>36.9$ (Fig.
  \ref{evolution}) and $2.42 \pm 0.05 < T_j < 2.79 \pm 0.09$. However, low inclination
  MMB is also possible, together with high inclination IMB.   

\begin{figure}[!!h]
\centering
\includegraphics[width=0.55\textwidth]{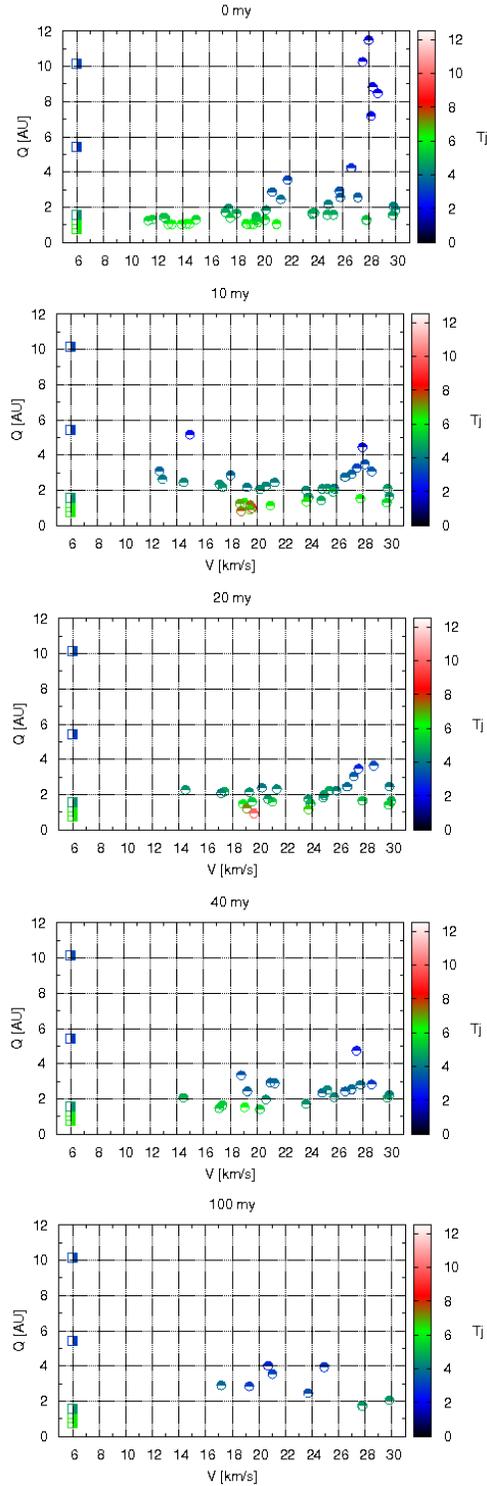}
\caption{Evolution of a sample, through the different ranges of admitted
  velocities, of fictitious asteroids (impactors) over the
  total integration time. In colours the $T_j$, x-axis is the impact velocity
 (km/s) and the y-axis is the aphelion (AU). On the left, at a fictitious 
velocity of 6 km/s we have the planets as reference, from top
to the bottom: Saturn, Jupiter, Mars, Earth, and Venus. The $T_j$ shows that the
particles tend to achieve the values of the MBAs (plot at 100 my).}
\label{one}
\end{figure} 

\clearpage

\begin{figure}[!h]
\centering
\includegraphics[width=0.8\textwidth]{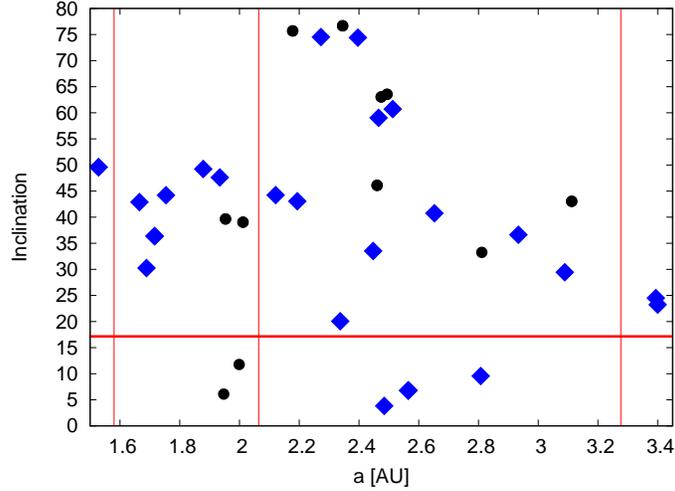}
\caption{ $a-i$ space. Vertical lines define the border of the regions in
  semi-major axis, as in Table \ref{region}, and the horizontal line discriminate \emph{HIG} and
  \emph{LIG}. Diamonds for \emph{GC} and circles for \emph{FC}.}
\label{evolution}
\end{figure} 

\end{description}

%\clearpage

\section{Conclusions}
The Bosumtwi impactor probably originated in the
MMB at orbital inclinations greater than $35^\circ$ with a possible initial
$T_j$ equal to $2.63\pm0.25$. These values are based only on our numerical
integrations (the highest
  values\footnote{Because of close encounters, some integrations were stopped before 100 My, so
less orbits end their evolution in the Main belt, see Fig. \ref{one}. These
ones are the missing percentage from the results, a part still happen to be
as NEAs, a part reach our maximum tolerance value for eccentricity (0.95) and
another small fraction has again impacts during the backward integration.} in
  $P_G(a,i)$ and $P_G(a)$ found in the \emph{GC}, see Table \ref{region} and
  \ref{MBAres} and Figure \ref{evolution}.)  and not considering the spectroscopical type too, because we do not have yet any significant number of measure in
this zone of the Main Belt (Cellino et al, 2002). Also this zone is still not
well studied and so we could not identify a particular family as the most likely source. Asteroids with similar orbital
parameters to the modeled Bosumtwi impactor are: 2002 MO$_3$, 2009 XF$_8$, 2002 SU and 2010 RR$_{30}$.
 There could be a cluster of asteroids at very high inclined orbits as shown by these results,
 something that we are planning to study after this work.
This method should be improved to find more consistent probabilities
(i.e. with more fictitious particles and larger integration times) and it
can potentially be applied to other old impact craters with well constrained
impactor properties, and even to impacts on other planets.

\noindent

\noindent

%%%%%%%%%%%%%%%%%%%%%%%%%%%%%%%%%%%%%%%%%%%%%%%%%%%%%%%%%%%%%%%%%%%%%%%
%%%%%%%%%%%%%%%%%%%%%%%%%%%%%%%%%%%%%%%%%%%%%%%%%%%%%%%%%%%%%%%%%%%%%%%

%\newpage
%% -------------------------------------

% -------------------------------------

% -------------------------------------


\begin{chapthebibliography}{1}


 \bibitem[Artemieva et al.(2004)]{Art2004} Artemieva, N., Karp, T., Milkereit, B.: 2004, GGG 5, 11016

\bibitem[Broz \& Vokrouhlicky(2008)]{Broz2008} Broz, M.; Vokrouhlicky, D.: 2008, MNRS 390, 715

\bibitem[Bus et al.(2004)]{Bin2004} Bus, S.~J., Binzel, R.~P., Volquardsen,  E.~L., Berghuis, J.~L.: 2004, BAAS 36, 1140

\bibitem[Bus \& Binzel(2002)]{Bus2002} Bus, S.~J., Binzel, R.~P.: 2002, Icarus 158, 146
%\bibitem[Cellino et al.(2002a)]{Cel2002a} Cellino, A., Zappala V. and Tedesco, E. ~F.: 2002, MPS 37, 1965

\bibitem[Cellino et al.(2002)b]{Cel2002}Cellino, A., Bus, S.~J. and Doressoundiram, A. and Lazzaro, D.: 2002, Asteroids III, 633

\bibitem[de L{\'e}on et al.(2010)]{Leo2010} de L{\'e}on,J.,Licandro,J. and 3 coauthors%,Serra-Ricart, M., Pinilla-Alonso, N., Campins, H.
:2010,A\&A 517,A23

\bibitem[Dvorak et al.(2004)]{Dvo2004} Dvorak, R., Pilat-Lohinger, E., Schwarz, R., Freistetter, F.: 2004, A\&A 426, L37

\bibitem[Eggl \& Dvorak(2010)]{Egg2010} Eggl, S., Dvorak, R.: 2010, LNP 790, 431

\bibitem[Fern{\'a}ndez et al.(2002)]{Fer2002} Fern{\'a}ndez,J.~A.,Gallardo,T.,Brunini,A.:2002,Icarus159,358

\bibitem[Fowler \& Chillemi(1992)]{Fow1992} Fowler, B., Chillemi, B.: 1992, MNRAS 423, 3074

\bibitem[Galiazzo et al.(2012)]{Gal2012} Galiazzo M.~A., Bazso, A., Dvorak R.:2012, PSS

\bibitem[Koeberl(1994)]{Koe1994} Koeberl, C.: 1994, GSA Special Paper 293, 133

\bibitem[Koeberl et al.(1997a)]{Koe1997a} Koeberl, C., Bottomley, R., Glass, B.~P.,  Storzer, D.: 1997a, GCA 61, 1745

\bibitem[Koeberl et al.(2007a)]{Koe2007a} Koeberl, C.; Milkereit, B., Overpeck, J. ~T., and 9 coauthors: 2007a, MPS 42, 483

\bibitem[Koeberl et al. (2007b)]{Koe2007b} Koeberl, C., Shukolyukov, A.,  Lugmair,G.~W., Guenter W.: 2007b, EPSL 256, 534

\bibitem[Jenniskens et al.(2010)]{Jen2010} Jenniskens,  P.,Vaubaillon,J.,Binzel, R.~P. and 13 coauthors: 2010, Met.Plan.Sc. 45, 1590

\bibitem[Neron de Surgy \& Laskar(1995)]{Las1995} Neron de Surgy, O., Laskar,  J: 1995, BAAS 27, 1172

\bibitem[Novakovi{\'c}, Cellino \&  Knezevi{\'c}(2011)]{Nov2011} Novakovi{\'c}, B., Cellino, A., Knezevi{\'c}, Z.: 2011,Icar. 216, 69

 \bibitem[Pravec et al.(2012)]{Pra2012} Pravec, P., Harris, A.~W., Kusnir{\'a}k, P. Gal{\'a}d, A. and Hornoch, K.: 2012, Icarus 221, 365

%\bibitem[Shaw \& Wasserburg]{Sha1982} Shaw, H.~F., Wasserburg, G.~J.: 1982,    EPSL 60, 155

\bibitem[Tsiganis et al.(2005)]{Tsi2005} Tsiganis,K., Varvoglis,H. and Dvorak,R.: 2005, CeMDA 71

\end{chapthebibliography}
\end{document}